\documentclass[pss]{wiley2sp} 
\usepackage{amsmath}
\usepackage{adjustbox}
\usepackage{color}

\begin{document}

\title{Tuning the work function of Si(100) surface by halogen absorption: a DFT study}

\titlerunning{Tuning the work function of Si(100) surface by halogen absorption: a DFT study}

\author{%
  Matteo Bertocchi\textsuperscript{\textsf{\bfseries 1}},
  Michele Amato\textsuperscript{\textsf{\bfseries 2}},
  Ivan Marri\textsuperscript{\textsf{\bfseries 3}},
  Stefano Ossicini\textsuperscript{\Ast,\textsf{\bfseries 1},\textsf{\bfseries 3}}}
\authorrunning{M. Bertocchi et al.}

\mail{e-mail
  \textsf{stefano.ossicini@unimore.it}, Phone:
  +39-0522-522211, Fax: +39-0522-522312}

\institute{%
  \textsuperscript{1}\,Dipartimento di Scienze e Metodi dell'Ingegneria,
             Universit\`{a} di Modena e Reggio Emilia,
             Via Amendola 2 Pad. Morselli, I-42122 Reggio Emilia, Italy\\
  \textsuperscript{2}\,Laboratoire de Physique des Solides and Centre de Nanosciences et de Nanotechnologies, CNRS, Univ. Paris-Sud,
             Universit\'{e} Paris-Saclay, 91405 Orsay, France\\
  \textsuperscript{3}\, CNR-Istituto di Nanoscienze-S3, via Campi 213 A, I-41125 Modena, Italy}


\abstract{%
%
%
%
\abstcol{%
First-principles calculations of work function tuning induced by different chemical terminations on Si(100) surface are presented and discussed. We find that
the presence of halogen atoms (I, Br, Cl, and F) leads to an increase of the work function if compared to the fully hydrogenated surface. This is a quite general effect and is directly linked to the chemisorbed atoms electronegativity as well as to the charge redistribution at the interface. All these results are examined with respect to previous theoretical works and experimental data obtained for the (100) as well as other Si surface orientations. Based on this analysis, we argue that the changes in the electronic properties caused by variations of the interfacial chemistry strongly depend on the chemisorbed species and much less on the surface crystal orientation.
}}

%
%
\titlefigure[width=\linewidth, scale=0.6 ]{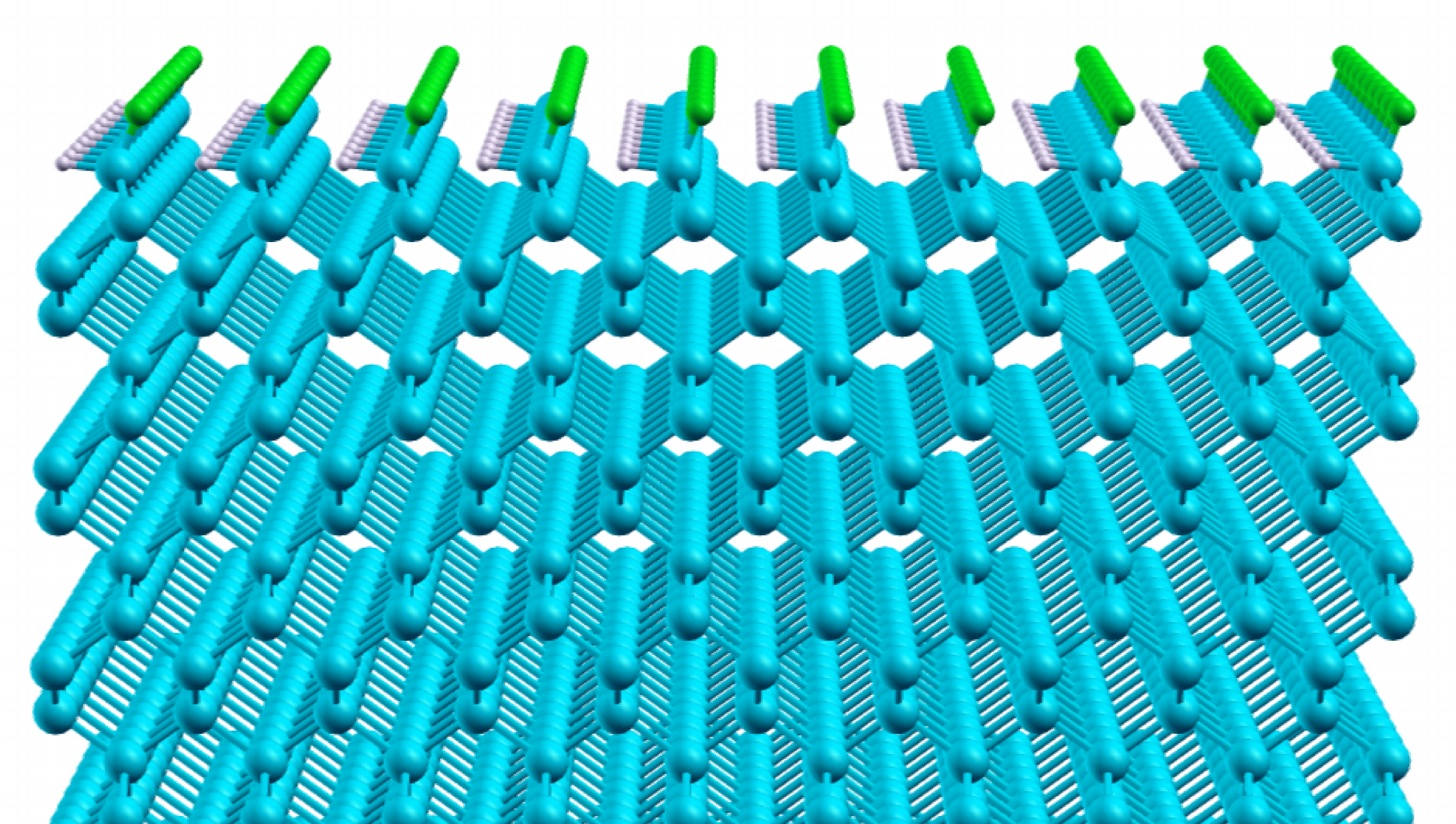}
\titlefigurecaption{%
A Si(100) surface supercell passivated by H and F atoms. Light blue spheres represent Si atoms, grey spheres H atoms and green spheres F atoms.}

%
%

\maketitle   

\section{Introduction}
\label{intro}
Silicon is one of the most used semiconductors. It is cheap, non-toxic and  largely engineered in electronic and micro-electronic devices for applications in different fields, from photonics to photovoltaics~\cite{SwansonPPRA2006,Pavesi2004}. In the last thirty years a large number of experimental and theoretical works have been dedicated to the study of the electronic and optical properties of Si-based systems with different dimensionality.\\
\indent In this context first principles techniques, based on the density functional theory (DFT), have emerged as efficient tools that can be used to complement experimental activities. The application of DFT, sometimes combined with advanced methodologies including many-body effects, has allowed to increase our knowledge about the structure, electronic, optical and transport properties of different semiconductor materials~\cite{BernardiPRL2014,Degoli_COMPTES,govoni_augerbulk}, to investigate how such properties are modified  by strain, passivation, and doping~\cite{iori_doping,guerra_NC} and to predict and quantify new effects, as the second-harmonic generation in silicon~\cite{SHG} and the carrier multiplication in isolated and interacting nanocrystals~\cite{govoni_nat,marri_JACS,Marri_Beilstein,MARRI_SOLMAT}.\\
\indent In this work we investigate the (100) surface of silicon. This system exhibits a variety of reconstructions that are mainly related to the different ordering of surface dimers. Since the Si(100) system is largely used in micro- and nano-electronic industry, strong efforts have been dedicated to the study of its chemical and physical properties ~\cite{DabrowskiBook2000,BuriakCR2002,AmatoPSS2016}.
From a broader perspective the study of the chemical passivation of group IV semiconductor surfaces is back to being a hot topic of debate. In particular, the analysis of the electronic modifications at the semiconductor surface induced by the adsorption/chemisorption of atoms and molecules has been the object of intense research activities. It has been pointed out that the functionalization of these surfaces is a key process in micro- and nano-electronics, energy conversion, charge storage, process information, sensing and electrochemical catalysis~\cite{BuriakCR2002,CummingsCCR2011,TeplyakovJVST2013,WongACR2014,AmatoCR2014,PengCR2015,FabreCR2016}. Indeed, chemisorption of atoms and molecules on semiconductor surfaces has a strong influence on some properties that are fundamental for technological applications, i.e. chemical reactivity, surface conductance, band edge profile and work function ($WF$). More specifically, the possibility to tune ionization potential, electron affinity and $WF$ of both the (111) and (100) surfaces of silicon~\cite{FabreCR2016,MorikawaPRB1995,SiciliaIJQC1997,BhaveJVST1996,DavydovTP2004,sgiarovelloPRB2001,KajitaJP2006,HeJACS2008,AnagawJPC2008,NovikovRJPC2010,NgTCA2010,CooperASS2011,KuoPCCP2011,HuangJAP2013,LiJPCC2013,ArefiJPCC2014,ArefiPCCP2016,NawazIEEE2015} has been investigated and exploited.\\
\indent The aim of this work is to study, within DFT, the effect induced by the chemisorption of halogen atoms on the electronic properties of the Si(100) surface. In particular, we aim at investigating, on an atomic scale, the mechanisms that are responsible of the $WF$ modulation. The accurate determination of the $WF$ provides an absolute electron energy level reference relative to the vacuum energy, which is important for applications like hetero-interfaces based devices, electrocatalysts, oxide and graphene based electronics, solar cells, Schottky junctions and thermionic devices.
The term of comparison throughout our work is the hydrogenated surface, which often represents the starting point for the preparation of samples with different chemical termination~\cite{HeJACS2008,LiJPCC2013}. In this work we consider only the unreconstructed (1x1) Si(100) surface. \\
\section{Theoretical Approach}
\label{theory}
Si(100) surface was modeled through the supercell method using a slab of $n$ semiconductor layers, grown along the (100) direction. In our calculations we adopt the bulk Si lattice parameter (a$_{Si~bulk}$=5.40~\AA). The determination of $n$ is generally done to satisfy specific requirements that are connected with  both the system and the physical properties under consideration; depending on them, $n$ can be tuned on a large range of values. The energy gap $E_G$, for instance, strongly depends on the slab thickness, as consequence of the quantum confinement effect~\cite{DelleyAPL1995,KholodSS2003,LiPRB2010}; a detailed determination of ${E_G}$ requires therefore an accurate convergence on $n$. The same considerations hold for the $WF$ that, in general, requires a large number of layers to be converged.
To avoid any ambiguity all the results presented in this work refer to a slab made of $n$ = 36 layers, that corresponds to a slab thickness  of about 50~\AA. By assuming $n$ = 36, we can correctly reproduce the electronic structure of bulk silicon thus ensuring an achieved convergence for all the calculated properties.\\
\indent In our simulations, Si slabs have been embedded in very large supercells, in order to avoid spurious interactions among periodic replicas that may influence the computed electronic properties. In particular, we have inserted a vacuum region of more than 35~\AA~\cite{AmatoJAP2016,NotesVac}. Surfaces have been modeled by adopting centrosymmetric slabs.
Since we considered an unreconstructed  Si(100) surface, both the top and the bottom surface layers of our slabs contain two-dangling bonds. Initially, all these dangling bonds have been saturated with hydrogen atoms. This hydrogenated structure represents our reference system, accordingly with the recipes often used in experiments (see Section~\ref{intro}). 
In order to understand how the properties of the Si surface depend on the atomic termination, we have investigated surfaces terminated with different adsorbed species, labeled by $X$. These species replace one of the initial capping H atoms placed at the top and bottom surface sites. The structures are then relaxed, keeping the atomic position of the central layers fixed, in order to reproduce a bulk-like region in the slab. In our investigation different $X$ terminations have been chosen by considering a series of halogen atoms (I, Br, Cl, and F) with different electronegativity. 
As previously done in several theoretical and experimental studies concerning the Si(111) surface~\cite{LiJPCC2013,ArefiJPCC2014}, we discuss our results in term of electronegativity and charge rearrangement at the surface.\\
\indent The calculations were performed employing the plane-wave pseudopotential PWscf code of the QuantumESPRESSO distribution~\cite{PWSCF2009}. This code is largely used by scientific community (and also by our group~\cite{marri_SSC,Iacomino_PRB,ossicini_NC,GUERRA_SM,iori_NC1}) and allows to calculate structural and electronic properties of metal and semiconductor materials using different exchange-correlation functionals. In our work we adopt the local density approximation (LDA) and we employ norm-conserving pseudopotential for all the chemical elements considered. As is known, in semiconductors, LDA severely underestimates the electronic band gap with respect to the experimental values. For the bulk silicon, for instance, we obtain $E_{G,bulk}^{Si}$ = 0.56 eV, to be compared to the experimental $E_{G,bulk,exp}^{Si}$ = 1.12 eV~\cite{AmatoJAP2016}. This underestimation can be, in principle, overcome by introducing quasiparticle corrections using the so-called GW approximation~\cite{HedinPR1965,OnidaREVMODPHYS2002}. This procedure is, however, quite demanding from a computational point of view. The accuracy of the method employed in this work relies on the fact that in several cases the use of the GW correction results in an almost rigid shift of the band, corresponding to a constant opening of the gap value. Moreover, as proven in previous works~\cite{LiJPCC2013}, the $WF$ can be accurately determined also in the framework of the DFT at the LDA level. Being the scope of this manuscript the study of the $WF$ in Si(100) surfaces with different passivations, we hence limit our approach to the DFT-LDA.\\
\indent Noticeably, the calculated values for the band gaps of Si surfaces with different passivation are very similar (within 40 meV) to the one obtained for the Si bulk. This result points out that, for the considered systems, passivation do not influence the energy gap and also that our slabs correctly describe the bulk properties of the surface.\\
\indent A careful analysis of the convergence of both the electronic and structural properties with respect to the plane-wave basis set cutoff has been conducted.  All the results discussed in the next sections concern the electronic structure calculated for the optimized geometries. 
\section{Electronic structure and work function}
\label{results_I}
\begin{figure}[h!]
\begin{center}
\includegraphics[width=0.55\textwidth]{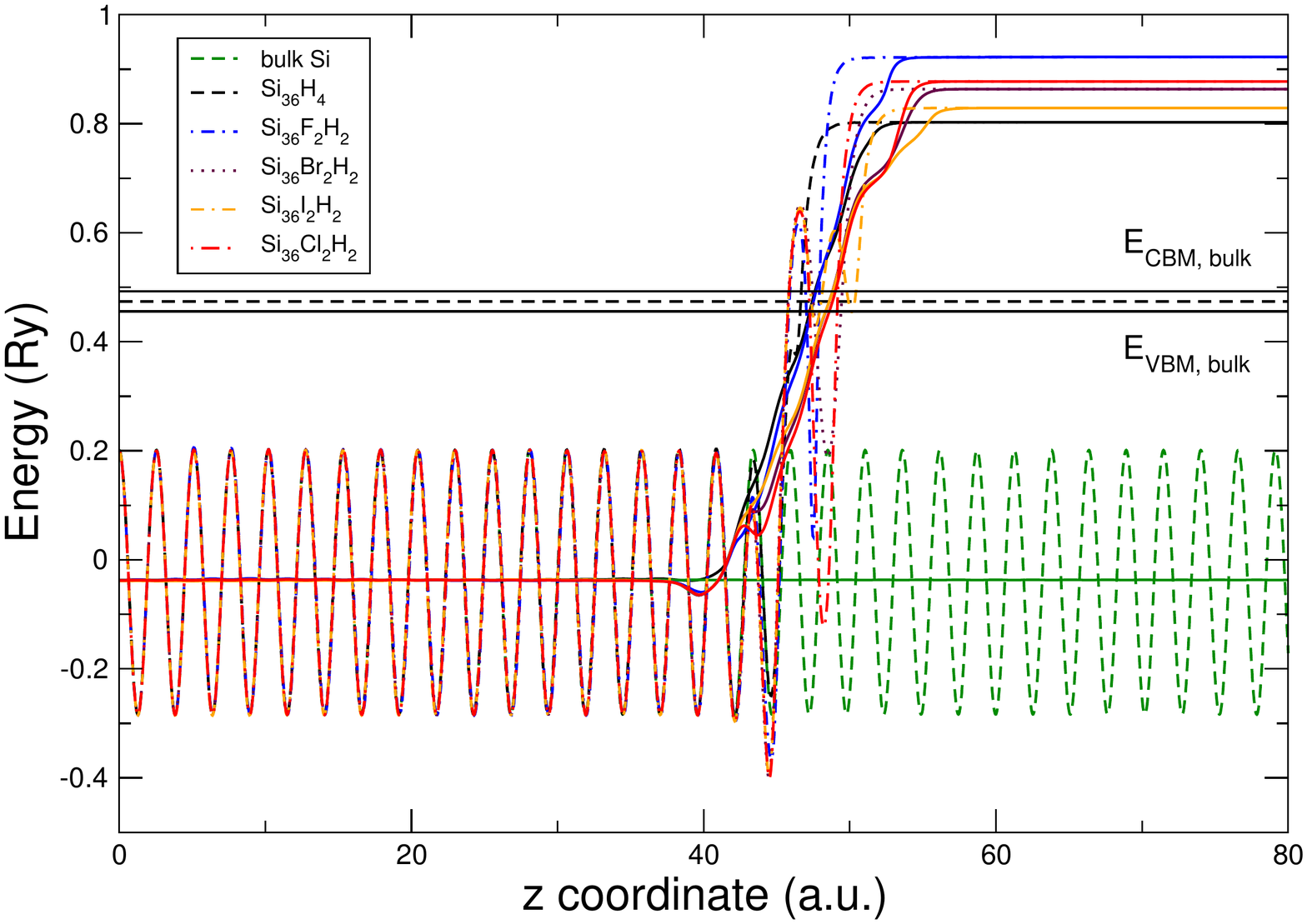}
\end{center}
\caption{(Color online) Planar (dashed lines) and macroscopic (solid lines) averages of the electrostatic potential for the different chemisorbed slabs. All the curves are superimposed in the bulk-like region of the slab to the oscillating planar average of the electrostatic potential calculated for the bulk Si supercell. The position in energy of the valence band maximum and conduction band minimum for the bulk Si supercell are highlighted. The Fermi level is taken at half of the bulk Si supercell energy gap. Energies are in Rydbergs.}
\label{Fig0}
\end{figure}

The $WF$ of a material is usually defined as the minimum energy required to remove an electron from the bulk to the vacuum outside across the surface. Consequently, it can be calculated as the difference between the vacuum energy level $E_{vac}$ and the Fermi energy $E_{F}$ of the system under consideration:
\begin{equation} 
WF = E_{vac} - E_{F}  
\end{equation}
In order to calculate the surface $WF$ by employing this equation, accurate values of both the vacuum potential and the Fermi energy are needed. In a semiconductor, since there are no allowed electronic energy levels in the gap, the Fermi energy is a somewhat theoretical construct. Moreover, in a vacuum-slab-supercell calculation, the $E_{F}$ can be calculated in different ways. 
For example, one can consider as Fermi energy the value directly derived from the slab calculation~\cite{AnagawJPC2008,MarzariPRB2009}. On the other hand, one could employ the Fermi energy of the corresponding bulk system whose value will not suffer from finite-size effect~\cite{ArefiPCCP2016,MarzariPRB2009,RusuJPCC2009}.
We have tested both the methods and proved that, for the considered systems, they give quite similar results if, in the first case, a slab formed by a sufficient number of layers and an adequate vacuum thicknesses are take into account.\\
\indent In this manuscript $WF$ is calculated considering the $E_{F}$ of bulk Si and using a procedure consisting of few simple steps~\cite{BorrielloPRB2007}: (i) firstly, starting from the unrelaxed passivated surface, we remove hydrogens, halogen atoms and the vacuum region, thus building a bulk Si supercell that contains the same number of Si atoms of the passivated surface. For this system we calculate the Fermi level $E_{F}$ taken at half of the energy band gap (see Fig.~\ref{Fig0}). (ii) Secondly, we consider H and $X$ passivated relaxed surface and we calculate electronic properties and the vacuum energy. The  vacuum level is determined by calculating the planar average of the electrostatic potential in the slab supercell (dashed lines in Fig.~\ref{Fig0}), far away from the surface, from which the macroscopic average along the $z$ direction is then deduced (solid lines in Fig.~\ref{Fig0}). (iii) As third step, we superimpose the oscillating planar average of the electrostatic potential calculated for the bulk Si supercell with the one calculated in the middle region of the passivated slab. We then calculate the $WF$ as the difference between the vacuum energy level and the $E_{F}$ of the bulk Si supercell~\cite{BorrielloPRB2007}. Results obtained are reported  in Fig.~\ref{Fig0}.
\label{results_III}

\begin{figure}[h!]
\vspace*{0.8cm}
\begin{center}
\includegraphics[width=0.55\textwidth]{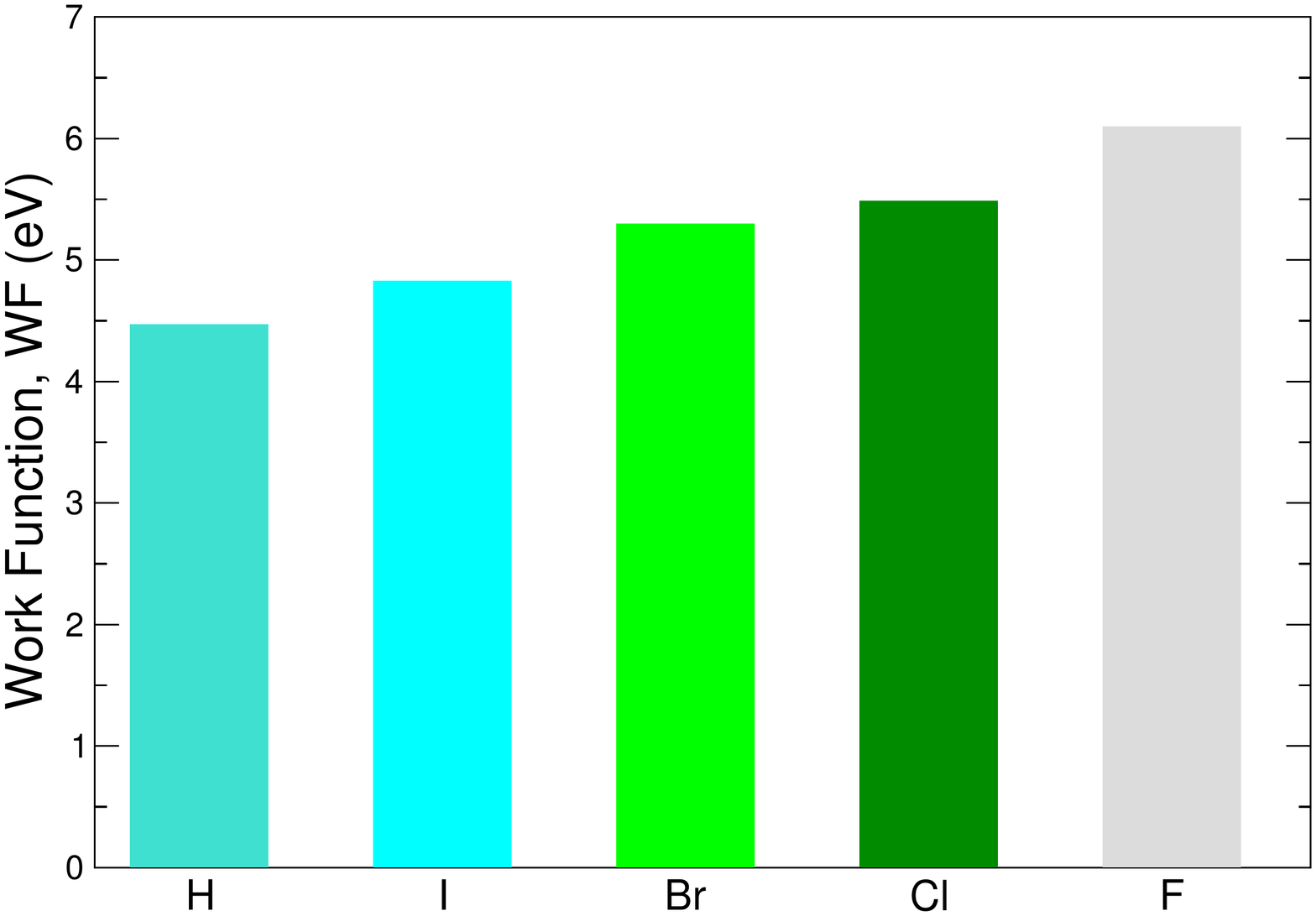}
\includegraphics[width=0.55\textwidth]{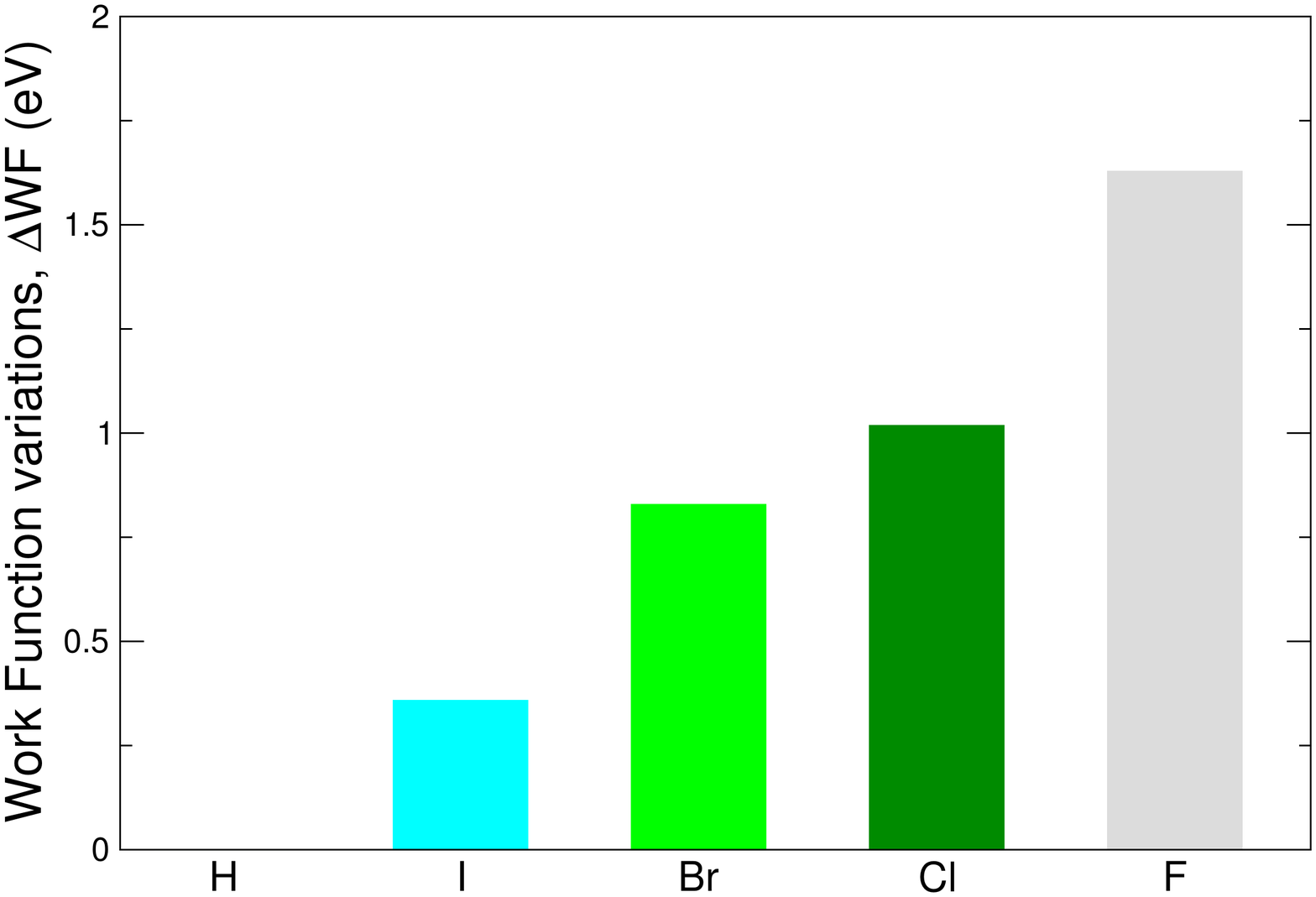}
\end{center}
\caption{(Color online) Top panel: Trends in the $WF$ values at DFT-LDA level (see text) for the Si(100) surfaces terminated by different atomic species. Bottom panel: Variations in the $WF$ values at DFT-LDA level (see text) for the Si(100) surfaces passivated with different atomic species with respect to the hydrogenated one.}
\label{Fig1}
\end{figure}

Table~\ref{TableI}~summarizes our calculated values of the $WF$ for the H, I, Br, Cl and F terminated Si(100) surface. This table also contains $\Delta WF$, i.e. the differences between the calculated $WF$ values for all the considered chemisorbed surfaces with respect to the hydrogenated ones.
\setlength{\tabcolsep}{10pt}
\begin{table}[h!]
\begin{center}
\caption{DFT-LDA calculated values (in eV) of the $WF$ for chemically modified Si(100) surface. The results (second column) are shown as function of $X$, the different atomic termination (first column). The third column reports the $WF$ changes, $\Delta WF$ (in eV), induced by the chemisorbed species with respect to the H-terminated surfaces.}
\begin{tabular}{@{}cccccccc@{}}
\hline
   $X$          &    $WF$  &  $\Delta WF$   \\
\hline
\hline
      H         &   4.47  &   0              \\
      I         &   4.83  &  +0.36           \\
      Br        &   5.30  &  +0.83           \\
      Cl        &   5.49  &  +1.02           \\
      F         &   6.10  &  +1.63        \\
\hline
\end{tabular}
\label{TableI}
\end{center}
\end{table}
Results of Table~\ref{TableI}, also reported in Fig.~\ref{Fig1} to improve the visibility of trends, point out that the calculated value for the hydrogen terminated surface is not so far from the value of the $WF$ calculated by Sgiarovello et al.~\cite{sgiarovelloPRB2001} for the (2x1) reconstructed Si(100) surface (5.12 eV), i.e. without hydrogen passivation. The agreement is noticeable if one takes into account the experimentally observed lowering (about 0.4 eV) of the $WF$ values of clean surfaces due to hydrogen adsorption~\cite{FukiwaraPRB1982,SouzisJVST1989}.

Fig.~\ref{Fig1} clearly indicates a similar trend for the different chemisorbed Si surfaces. All the considered halogen passivated surfaces, indeed, show an increment of $WF$ with respect to the H-terminated one.\\
\indent The comparison between the calculated values of $WF$ and the experimental results is not trivial due to several reasons. Firstly, the methodologies adopted to measure the $WF$ can sometimes lead to different results. It is worth mentioning, for instance, that the $WF$ can be measured both by using photoelectron spectroscopy and Kelvin probe technique~\cite{PouchJPCC2015}. The first method allows the measurement of the absolute $WF$, whereas the second gives only a contact potential difference between the probe and the sample surface. However, using a calibration procedure with photoelectron spectroscopy, it is possible to transform the Kelvin probe results into absolute values, thus permitting a more uniform interpretation of the experimental data. Secondly, experimental values of the $WF$ are often influenced by the characteristic of the sample, like the presence of defects, doping and impurities concentration, and so on. These parameters affect the experimental results but cannot be easily included in a numerical ab-initio model. Finally, the intrinsic approximations present in the methodology adopted for the calculation of $WF$ (in our case the DFT-LDA approximation) can lead to an over-estimation or to a under-estimation of the experimental data.\\
\indent Experimental results for the $WF$ of the Si(100)2x1 clean surface range from 4.6 eV to 4.91 eV~\cite{KonoASS1989,GunsterSS1996,LuPRB1996}. These values are larger (with differences from 0.13 eV to 0.44 eV) than our 4.47 eV calculated for the H-terminated Si(100)1x1 surface. We have to consider, however,  the experimentally measured decrease of the $WF$ on going from the clean Si(100)2x1 to the H-Si(100)2x1 (from 0.34 eV to 0.40 eV)~\cite{FukiwaraPRB1982,SouzisJVST1989}. Additionally, He et al.~\cite{HeJACS2008}, by employing photoelectron spectroscopy, reported $WF$ values of H-passivated Si(100) surfaces (for different doping concentration) in the range between 4.21 eV to 4.55 eV, values that go along with our result of 4.47 eV.
\setlength{\tabcolsep}{8pt}
\begin{table}[b!]
\begin{center}
\caption{$WF$ values calculated for the differently terminated Si(100) surfaces. The value in parenthesis are related to the $WF$ changes, $\Delta WF$, induced by the different chemisorbed species with respect to the H-terminated surfaces. Our results for the Si(100) surfaces are compared with theoretical~\cite{ArefiJPCC2014} and experimental results~\cite{LiJPCC2013,HackerSSE2010,HungerJAP2002,HungerPRB2005,HungerJPC2006,LopinskiPRB2005} obtained for the Si(111) surface. All the values are in eV.}
\begin{tabular}{@{}ccccc@{}}
\hline
   $X$  &    Present Work  &  Th.(50$\%$)~\cite{ArefiJPCC2014}&Th.(100$\%$)~\cite{ArefiJPCC2014}\\
\hline
\hline
      H         &   4.47~(0.0)  &  4.35~(0.0)    &4.35~(0.0)  \\
      I         &   4.83~(+0.36)  & 4.72~(+0.37)  & 4.62~(+0.27)\\
      Br        &   5.30~(+0.83)  & 4.98~(+0.63)  & 5.12~(+0.77)\\
      Cl        &   5.49~(+1.02)  &  5.09~(+0.74) & 5.43~(+1.12)\\
      F         &   6.10~(+1.63)  &  5.36~(+1.01) & 6.39~(+2.04)\\
\hline
\hline
   $X$  &Exp.~\cite{LiJPCC2013}&Exp.~\cite{HackerSSE2010}&Exp.~\cite{HungerJAP2002,HungerPRB2005,HungerJPC2006,LopinskiPRB2005}\\
\hline
\hline
      H         & 4.16-4.24~(0.0)&4.17-4.35~(0.0) &4.42(0.0)\\
      I        &              &            &\\
      Br        & 4.32~(+0.16) &           &\\
      Cl        &  4.60~(+0.36) &           & (+1.12-1.15) \\
      F         &               &           & \\
\hline
\end{tabular}
\label{TableII}
\end{center}
\end{table}

It also possible to compare, in some cases, our results with other theoretical calculations. Anagaw et al.~\cite{AnagawJPC2008} have investigated, using DFT, the change in the $WF$ induced by chemisorption on Si(100)2x1 surfaces. They found 4.6 eV for the H-terminated surface that is in good agreement with our results for the Si(100)1x1 surface (4.47 eV). Ng et al.~\cite{NgTCA2010} have investigated the modulation of the $WF$ of silicon nanowires (NWs) grown in the (110) direction, whose dangling bonds at the interface are firstly covered by hydrogens that are then substituted with OH or F. For NWs with diameter of about 1 nm they found $WF$ values of 4.61 eV for the fully H-covered system and 5.31 eV~-~6.73 eV for the F-terminated case, depending on the percentage of coverage. This behaviour, that consists of an increase of the $WF$ for the F-covered NWs, is congruent with what we have found for F terminated surfaces.  
\begin{figure}[b!]
\begin{center}
\includegraphics[width=0.55\textwidth]{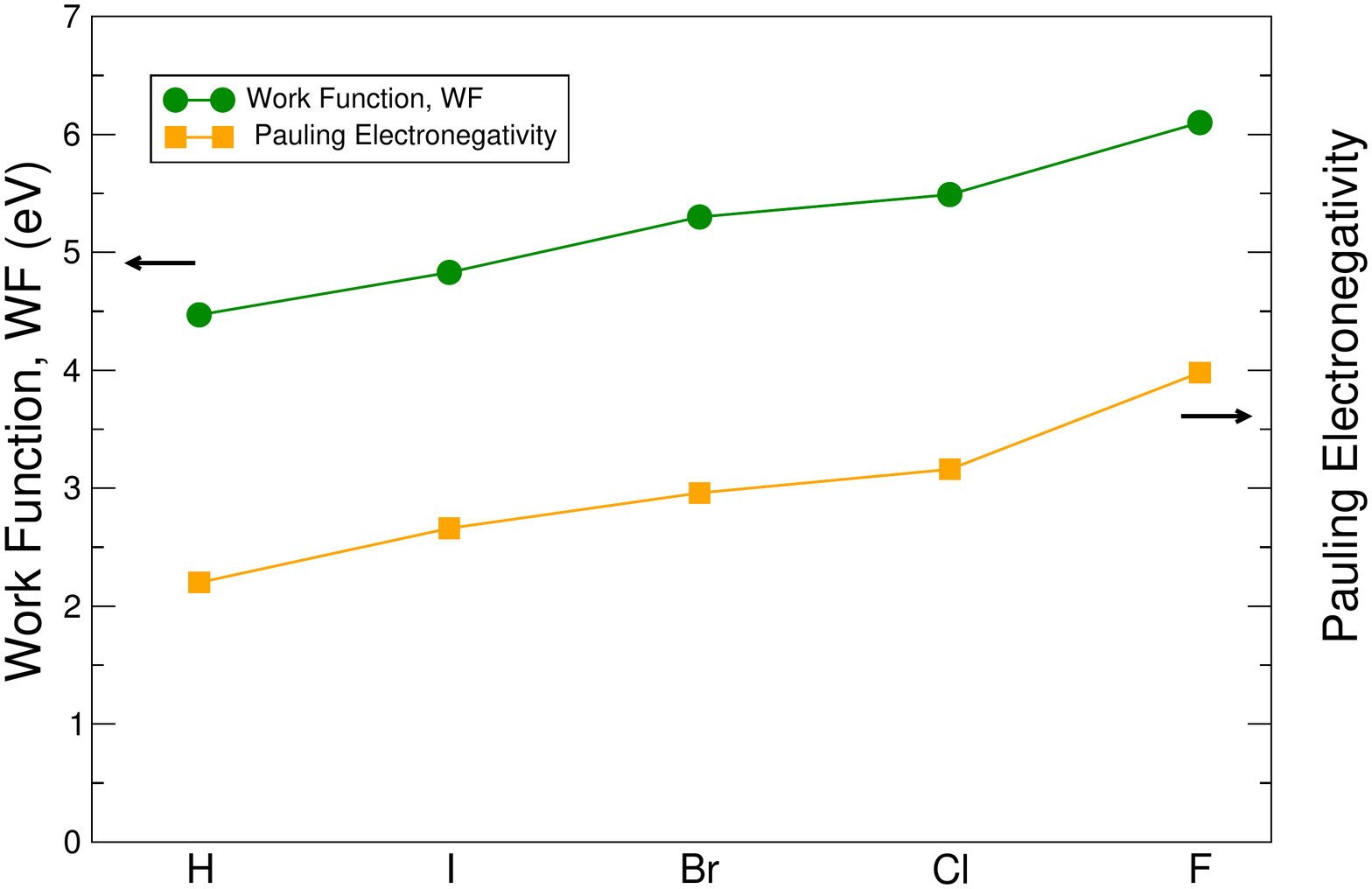}
\end{center}
\caption{(Color online) Calculated $WF$ values (in eV) for Si(100) surfaces with different atomic termination (green dots). Orange squares represent the Pauling electronegativity of the chemisorbed atoms. The lines are guide for the eyes.}
\label{Fig3}
\end{figure} 

In Table~\ref{TableII} we correlate our calculated $WF$ with theoretical and experimental results obtained for the Si(111) surfaces~\cite{LiJPCC2013,ArefiJPCC2014,HackerSSE2010,HungerJAP2002,HungerPRB2005,HungerJPC2006,LopinskiPRB2005}. Noticeably, in Ref.~\cite{ArefiJPCC2014}, the $WF$ has been calculated for two different coverages of the Si(111) surface (50$\%$ and 100$\%$): the first one is the same of our Si(100) surface, where only one of the two capping H atoms has been substituted by other species. We see that the agreement is indeed quite good. Moreover Li et al.~\cite{LiJPCC2013} have calculated at G$_0$W$_0$ level
the variation in the ionization potential with respect to a H-terminated Si(111) surface due to the substitution of H with Br and Cl. They found variation of 0.8 and 1.1 eV to be compared to our 0.83 and 1.02 eV values, respectively. Since the variations in the ionization potential cannot be very different from those of the $WF$, these results confirm the accuracy of LDA in evaluating the $WF$ for the considered systems. The agreement between our calculated $WF$ changes for the Si(100) surfaces with those determined for Si(111) surfaces is quite remarkable: we can observe a very similar increase of the $WF$ in all the halogen passivated surfaces. These results are a strong indication that, for this quantity, the major role is played by the chemisorbed species present on the surfaces and not by the different surface orientation.\\
\indent As previously stated, the dependence of $WF$ values on the passivating species for the Si surface (see Fig.~\ref{Fig3}) shows that halogen atom chemisorption (I, Br, Cl and F) always induces an increase in the $WF$.
With the purpose of finding a rationale for the observed trend, in Fig.~\ref{Fig3} we also reported the Pauling electronegativity values for each halogen atom. 
It is clear that the augmentation of the $WF$ value for the passivated Si surface follows the same trend for the increase of the electronegativity going from I to F. Since the Pauling electronegativity of Si is 1.9, smaller than those of all the atoms presented in Fig. 3, electrons are attracted to the passivating layer in a manner directly linked to the relative differences between the electronegativity of Si and that of the chemisorbed species, thus determining a dipole moment at the adsorbate/substrate interface~\cite{LiJPCC2013}. This dipole moment, that mainly depends on the local chemistry between the adsorbate and silicon and much less on the surface orientation, is responsible of the variations in the work function, that results, hence, quite independent on the atomic arrangement of the Si surface.
\begin{table}[h!]
\begin{center}
\caption{$\Delta_q(Si)$ indicates the calculated L\"{o}wdin charge change for the Si atom in contact with the halogen element, while $q_x$ is the calculated partial charge of halogen atom. All the values are in units of electron charge. Theoretical results for the Si(111) surface~\cite{ArefiJPCC2014} are also reported.}
\begin{tabular}{@{}cccccccc@{}}
\hline
   $X$          &    $\Delta_q(Si)$  & $q_x$ & $\Delta_q(Si)$~\cite{ArefiJPCC2014} & $q_x$~\cite{ArefiJPCC2014}  \\
\hline
\hline
      I         &  -0.04   &     -7.02   & -0.03 & -7.07 \\
      Br        &  -0.17   &      -7.12  & -0.13 & -7.16 \\
      Cl        &  -0.25   &      -7.18  & -0.22 & -7.22 \\
      F         &  -0.53   &      -7.36  & -0.53 & -7.46 \\
\hline
\end{tabular}
\label{TableIII}
\end{center}
\end{table}

In Table~\ref{TableIII} we report the L\"{o}wdin charge changes $\Delta_q(Si)$ calculated for the Si atom in contact with the halogen element with respect the fully hydrogenated configuration together with $q_x$ the computed partial charge of the halogen atoms. The data, in agreement with results of Ref.~\cite{ArefiJPCC2014} for the (111) Si surface, indicates an increment of $\Delta_q(Si)$ when we move from I to F meaning that the charge transfer is more pronounced when the halogen electronegativity increases, reflected also in the augmentation of $q_x$. These results point out the direct relation that exists between $\Delta WF$ and the redistribution of charge at the surface. Furthermore, this confirms again the dominant role of the chemisorbed species with respect to the surface orientation in the determination of the $WF$. 
\section{Conclusions}
We presented and discussed first principles simulations of halogen passivated Si(100) surface. By calculating work function and charge transfer we were able to draw up a general trend for the halogen absorption effect on the Si electronic properties. In particular, we have found that halogens have a tendency to increase the $WF$ of Si surfaces as a consequence of their high electronegativity that causes a large charge transfer with an associated dipole moment. The behavior observed moving from I to F fairly follows the increase in the corresponding Pauling electronegativity. The generality of this effect is confirmed by the discussion and comparison of our results with other calculations and experiments on both Si(100) and Si(111). This sheds light on the major role of the type of passivation with respect to the surface crystal orientation.  
\begin{acknowledgement}
M.A. is grateful to G. Cantele for useful discussions. This work was made possible thanks to the HPC resources of IDRIS under the allocation i2015097422 made available by GENCI (Grand Equipement National de Calcul Intensif) and the computer resources, technical expertise, and assistance provided by the Red Espa\~nola de Supercomputaci\'on. 
The authors thank the Super-Computing Interuniversity Consortium CINECA for support and high-performance computing resources under the Italian Super-Computing Resource Allocation (ISCRA) initiative, PRACE for awarding us access to resource FERMI IBM BGQ, and MARCONI HPC cluster based in Italy at CINECA. Ivan Marri acknowledges support/funding from European Union H2020-EINFRA-2015-1 programme under grant agreement No. 676598 project “MaX − materials at the exascale”.
\end{acknowledgement}

\bibliographystyle{pss}
\bibliography{manuscript}
\end{document}